\documentclass{article} 
\usepackage{iclr2026_conference,times}

\usepackage{amsmath,amsfonts,bm}

\def\eqref#1{equation~\ref{#1}}

\def\1{\bm{1}}

\DeclareMathAlphabet{\mathsfit}{\encodingdefault}{\sfdefault}{m}{sl}
\SetMathAlphabet{\mathsfit}{bold}{\encodingdefault}{\sfdefault}{bx}{n}

\usepackage{hyperref}
\usepackage{booktabs}
\usepackage{multirow}
\usepackage{url}
\usepackage{graphicx}

\usepackage{tabularx}
\usepackage{booktabs}
\usepackage{multirow}
\usepackage{makecell} 

\title{Auditing Cascading Risks in Multi-Agent Systems via Semantic–Geometric Co-evolution\thanks{This work has been accepted to ICLR 2026 Workshop: Principled Design for Trustworthy AI}}

\author{
Zixun Luo \\
Huazhong University of Science and Technology \\
\texttt{luozixun@hust.edu.cn} \\
\And
Yuhang Fan \\
Lingnan University \\
\texttt{fanyuhang@ln.hk} \\
\And
Hengyu Lin \\
Tsinghua University \\
\texttt{linhengyu@tsinghua.edu.cn} \\
\And
Yufei Li \\
Centre for Artificial Intelligence and Robotics \\
Hong Kong Institute of Science and Innovation \\
Chinese Academy of Sciences \\
\texttt{yufei\_li19972@outlook.com} \\
\And
Youzhi Zhang \\
Centre for Artificial Intelligence and Robotics \\
Hong Kong Institute of Science and Innovation \\
Chinese Academy of Sciences \\
\texttt{youzhi.zhang@cair-cas.org.hk}
}

\iclrfinalcopy 
\begin{document}

\maketitle

\begin{abstract}
Large Language model (LLM)-based Multi-Agent Systems (MAS) are prone to cascading risks, where early-stage interactions remain semantically fluent and policy-compliant, yet the underlying interaction dynamics begin to distort in ways that amplify latent instability or misalignment. Traditional auditing methods that focus on per-message semantic content are inherently reactive and lagging, failing to capture these early structural precursors.
In this paper, we propose a principled framework for cascading-risk detection grounded in semantic–geometric co-evolution. We model MAS interactions as dynamic graphs and introduce Ollivier–Ricci Curvature (ORC)—a discrete geometric measure—to characterize information redundancy and bottleneck formation in communication topologies. By coupling semantic flow signals with graph geometry, the framework learns the normal co-evolutionary dynamics of trusted collaboration and treats deviations from this coupled manifold as early-warning signals.
Experiments on a suite of cascading-risk scenarios aligned with the risk category demonstrate that curvature anomalies systematically precede explicit semantic violations by several interaction turns, enabling proactive intervention. Furthermore, the local nature of Ricci curvature provides principled interpretability for root-cause attribution, identifying specific agents or links that precipitate the collapse of trustworthy collaboration.
\end{abstract}

\section{Introduction}
The paradigm of Large Language Model (LLM)-based Multi-Agent Systems (MAS) has transitioned AI from static query–response loops to complex, self-evolving collaboration topologies \citep{GenerativeAgents}. While such systems excel at decomposing and coordinating intricate tasks, they introduce a critical and underexplored security challenge: cascading risk \citep{zhu2025llm}. Unlike atomic “jailbreaks” in single models \citep{yi2024jailbreak,wei2023jailbroken}, failures in MAS—such as collusion, hallucination cascades, or role misalignment—are often emergent properties of interaction dynamics rather than isolated semantic violations. In these scenarios, early-stage messages typically remain fluent and superficially policy-compliant, yet the underlying interaction structure gradually distorts, amplifying latent instability or harmful intent.

Current safety paradigms predominantly rely on semantic auditing, monitoring textual content for explicit violations \citep{zhuo2023red}. However, as MAS interactions scale in length and complexity \citep{li2024survey}, such approaches are inherently reactive \citep{ray2019benchmarking,amodei2016concrete}. By the time a semantic violation becomes observable, the collaborative dynamics of the system have often already collapsed into an irreversible failure mode. This reveals a fundamental detection gap between the onset of structural instability in Multi-Agent interaction and its eventual semantic manifestation \citep{peng2024jailbreaking}.


In this work, we propose a principled shift from reactive content filtering to proactive structural auditing. We posit that trustworthy MAS collaboration can be modeled as trajectories constrained by a stable semantic--geometric manifold \citep{marks2023geometry}. Analogous to physical systems that exhibit stress accumulation before catastrophic failure \citep{helbing2013globally}, MAS interactions display topological distortions—such as abnormal information bottlenecks \citep{tishby2015deep} or excessive redundancy—well before semantic content turns explicitly harmful. Capturing these structural precursors requires modeling not only what agents say, but how semantic information propagates through the interaction topology \citep{burdalo2018information} over time.

Existing research has explored graph-based anomalies in MAS communication networks to identify malicious nodes \citep{ma2021comprehensive}, and leveraged LLM-based evaluators to filter point-wise semantic violations \citep{gu2024survey}. However, these methods remain decoupled: they treat the interaction topology as a static container and the semantic content as an isolated stream. They fail to account for the dynamic coupling between what is being communicated and how the underlying interaction geometry deforms in response to that information. Consequently, subtle precursors of systemic collapse—where the semantic weight of an interaction outpaces the structural capacity of the network—remain undetected.

\begin{figure}[t]
    \centering
    \includegraphics[width=0.8\linewidth]{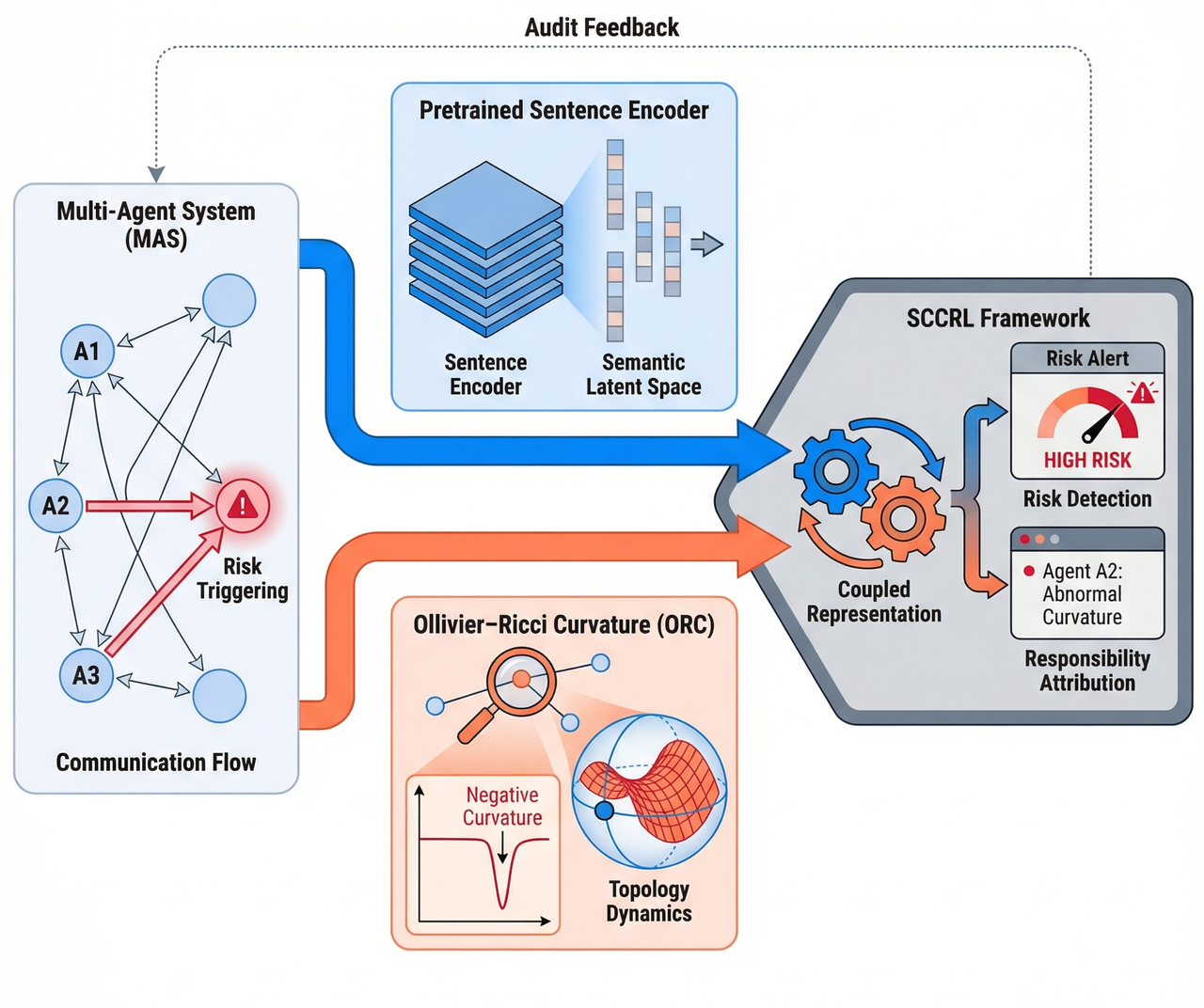}
    \caption{\textbf{Architecture of the Semantic-Curvature Co-evolution Auditing Loop (SCCAL).} 
    The framework treats multi-agent interactions as a dynamic graph stream $\mathcal{G}_t = (\mathcal{V}, \mathcal{E}_t)$ and operates via three integrated stages: 
    (1) \textbf{Dual-View Feature Extraction:} Interaction data is decoupled into \textit{Semantic Content} (node-level latent manifolds $\mathcal{M}_s$ via frozen encoders) and \textit{Interaction Topology} (edge-level geometry). 
    (2) \textbf{Geometric Quantization:} We utilize Ollivier-Ricci Curvature (ORC) to identify structural patterns; positive curvature indicates \textit{information redundancy} (potential echo chambers), while negative curvature highlights \textit{structural bottlenecks} (vulnerability for cascading risks). 
    (3) \textbf{Proactive Audit Loop:} By monitoring \textit{Consistency Residuals} between semantic and geometric flows, the system triggers \textbf{Audit Feedback}—a control signal that serves as an early-warning intervention to steer agents away from identified geometric traps before explicit semantic violations occur.}
    \label{fig:intro}
\end{figure}

To this end, we introduce the Semantic–Curvature Co-evolutionary Auditing Loop (SCCAL). As shown in Figure \ref{fig:intro}, our framework couples semantic flows—modeled via transmissibility and credibility—with interaction geometry quantified by Ollivier–Ricci Curvature (ORC), a discrete measure of local redundancy and fragility in information transport. SCCAL learns the normal co-evolutionary dynamics of semantic states and graph geometry from benign trajectories, and formalizes risk as a consistency residual: a deviation from the predicted semantic–geometric evolution path of the system. This formulation enables early detection of cascading instability, even when individual agent messages remain fluent, aligned, and non-violating.

Our contributions are threefold:
\begin{itemize}
    \item \textbf{A Fundamental Auditing Framework.} We formalize MAS safety as a trajectory stability problem on a coupled semantic–geometric manifold, providing a mechanistic and model-agnostic foundation for auditing cascading risk beyond black-box semantic classification.
    \item \textbf{Proactive Early Warning via Co-evolutionary Inconsistency}. We demonstrate that semantic–geometric inconsistency, rather than semantic signals or curvature alone, systematically emerges several interaction rounds before semantic-only baselines can detect risk, across both adversarial and attack-free failure regimes.
    \item \textbf{Interpretability and Root-Cause Attribution.} We show that curvature patterns—such as excessive positive curvature associated with redundant amplification and negative curvature associated with fragile information flow—offer interpretable structural signals for localizing the agents or interaction links that initiate cascading system failure.
\end{itemize}

\section{Related Work}
\paragraph{Safety and Robustness in Multi-Agent Systems.}

 Traditional safety analysis in Multi-Agent systems (MAS) focuses on strategic robustness within defined action spaces \citep{gleave2019adversarial}. While effective for closed-loop control, these methods struggle with the emergent risks in LLM-based systems \citep{GenerativeAgents}. Recent studies have highlighted that LLM-based agents are susceptible to semantic misalignments and adversarial social engineering \citep{yullm, perez2022red}, where harmful behaviors emerge not from code errors but from complex linguistic reasoning. Furthermore, risks in LLM-MAS often manifest as hallucination cascades \citep{huang2025survey} and collaborative toxicity \citep{deshpande2023toxicity}, propagating implicitly through semantic contagion and feedback loops. Unlike rigid policy execution, risks in LLM-MAS propagate implicitly through semantic contagion and feedback loops. This shift renders conventional constraint-based verification insufficient, necessitating novel auditing frameworks that can capture the interplay between unconstrained linguistic interaction and systemic risk propagation.

\paragraph{Semantic Safety and Adversarial Attacks.}
 Current LLM safety research predominantly targets atomic semantic failures—such as toxicity, hallucinations, and jailbreaks \citep{zou2023universal, yi2024jailbreak}, relying on latent space classification or auxiliary guardrails \citep{wei2023jailbroken, wang2023decodingtrust}. However, these approaches typically treat risk as an isolated event within a single turn. They overlook cascading risks in collaborative environments, where structural dependencies can amplify subtle semantic triggers that remain invisible to independent semantic audits.

\paragraph{Graph Geometry and Structural Anomaly Detection.}
Graph Neural Networks (GNNs) have been widely applied to detect structural anomalies \citep{ma2021comprehensive}, yet purely topological metrics often fail to distinguish benign dense collaboration from malicious collusion in MAS \citep{akoglu2015graph}. Recently, Ollivier-Ricci Curvature (ORC) has emerged as a powerful tool for characterizing the geometry of information flow, with frameworks like CurvGAD \citep{grover2025curvgad} demonstrating utility in distinguishing structural bottlenecks from background noise. However, existing curvature-based methods are largely static and agnostic to semantic context. In LLM-based systems, must be grounded in semantic dynamics to avoid false positives arising from healthy coordination.

\section{Methodology: Semantic–Geometric Co-evolutionary Auditing}\label{sec:method}

\begin{figure*}[t] 
    \centering
    \includegraphics[width=0.8\textwidth]{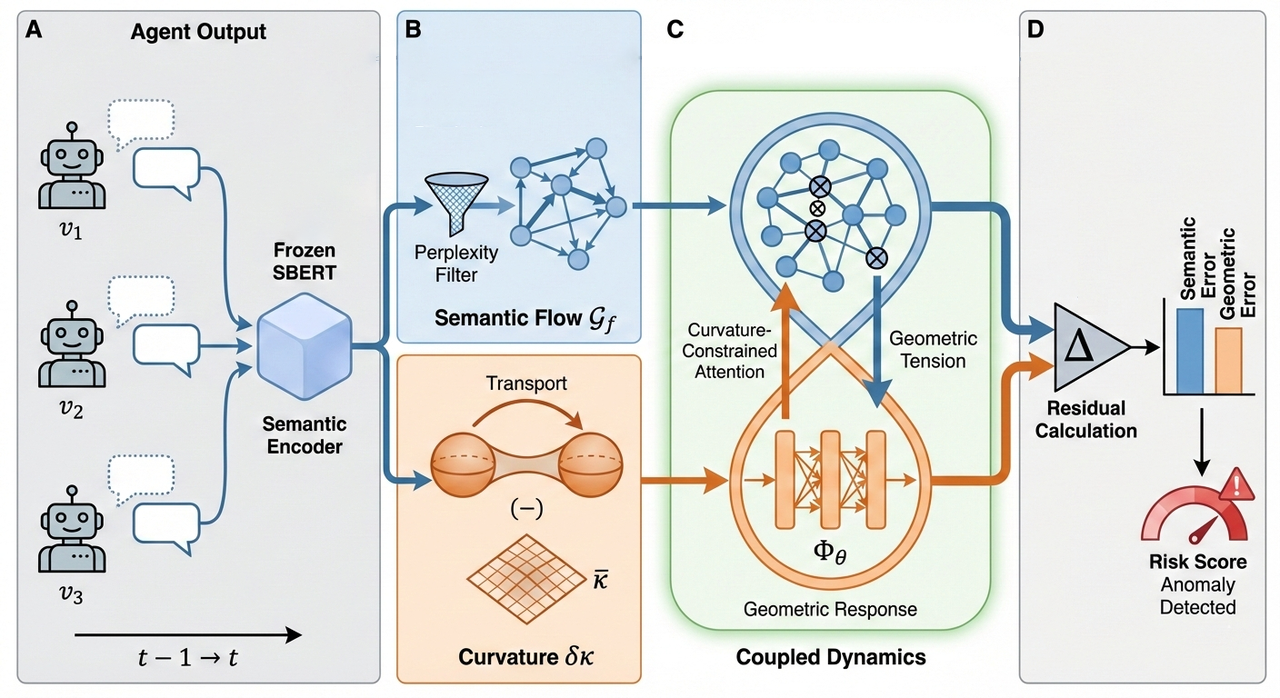} 
    \caption{\textbf{Overview of the proposed Semantic-Curvature Co-Evolution Loop Framework.} The framework models Multi-Agent interaction risks through four distinct stages: 
    \textbf{(A) Semantic Encoding:} Agent-generated messages are mapped into a latent space using a frozen SBERT to decouple representation from dynamics. 
    \textbf{(B) Semantic Flow Induction:} A weighted interaction graph $G_f$ is constructed based on semantic transmissibility and a Perplexity Filter. 
    \textbf{(C) Coupled Dynamics Modeling:} The core module $\Phi_\theta$ jointly models the bidirectional constraints between semantic updates and local topological evolution. Here, \textbf{Curvature $\bar{\kappa}$ (Orange Box)} represents the aggregated Ollivier-Ricci Curvature across a sliding interaction window, serving as a geometric descriptor of structural stability. 
    \textbf{(D) Anomaly Detection:} Risk signals are identified by the joint prediction residual. The \textbf{Geometric Response (Green Box)} quantifies the system's structural reaction to semantic shifts; a high deviation between the predicted and observed Geometric Response indicates a consistency violation, flagging potential cascading risks before they manifest semantically.}
    \label{fig:p1}
\end{figure*}

We conceptualize safety auditing in LLM-based Multi-Agent Systems (MAS) as a \emph{trajectory stability problem} on a coupled semantic–geometric manifold.
Rather than detecting risks through reactive content filtering, our framework, \textbf{SCCAL}, models the internal structural stress induced by semantic interactions and identifies cascading risks as violations of semantic–geometric consistency.
Figure~\ref{fig:p1} provides an overview.

\subsection{Auditing Manifold and Co-evolution Assumption}

A MAS interaction trace is represented as a sequence $\mathcal{S}=\{S_1,\dots,S_T\}$, where each system state
$
S_t=(\mathcal{G}_t,\mathbf{Y}_t)
$
consists of a directed semantic flow graph $\mathcal{G}_t=(\mathcal{V},\mathcal{E}_t,\mathbf{W}_t)$ and agent semantic states $\mathbf{Y}_t\in\mathbb{R}^{N\times d}$.

\paragraph{Semantic Representation.}
Each agent message $s_i^t$ is mapped into a latent semantic space via a frozen encoder $\mathcal{R}$:
$
\mathbf{y}_i^t=\mathcal{R}(s_i^t),
$
which decouples semantic representation from dynamics learning and enforces a stable semantic reference frame.

\paragraph{Principle of Co-evolution.}
In a trustworthy MAS, semantic evolution and interaction topology are interdependent rather than independent.
We model this dependency as a coupled dynamical system:
\[
\begin{cases}
\mathbf{Y}_{t+1}=\psi(\mathbf{Y}_t,\mathcal{G}_t;\theta_y)+\epsilon_y,\\
\mathcal{G}_{t+1}=\phi(\mathcal{G}_t,\mathbf{Y}_t;\theta_g)+\epsilon_g,
\end{cases}
\]
where $\epsilon$ denotes bounded natural variation.
Cascading risks correspond to \emph{consistency violations}, i.e., system states that deviate from the learned co-evolutionary manifold even when local semantic outputs remain coherent.

Importantly, neither semantic nor geometric dynamics alone admit a stable predictor under cascading risk, making joint co-evolution a necessary condition for consistent auditing.

\subsection{Directed Semantic Flow Induction}

Drawing inspiration from \citep{wang2023decodingtrust} and \citep{lian2025semantic}, we project the message passing process of the MAS into a latent space.

To capture causal semantic influence, we induce a weighted directed interaction graph from raw language exchanges.
Each message $s_i^t$ is encoded as $\mathbf{y}_i^t$, and the semantic flow from agent $i$ to agent $j$ is defined as:
\[
w_{ij}^t = \tau_{ij}^t \cdot \chi_i^t,
\]
where semantic transmissibility
$
\tau_{ij}^t=\cos(\mathbf{y}_i^t,\mathbf{y}_j^{t-1})
$
measures alignment between intents, and credibility
$
\chi_i^t=\exp(-\mathrm{PPL}(s_i^t))
$
penalizes incoherent or hallucinated outputs using a reference language model.
This construction suppresses spurious semantic propagation while preserving meaningful influence paths.

\subsection{Geometric Profiling via Ollivier–Ricci Curvature}

Inspired by \citep{grover2025curvgad}, we apply the Ollivier-Ricci Curvature (ORC) to the computation of the evolution process of the MAS interaction network.

We characterize the structural response of the interaction network using Ollivier–Ricci Curvature (ORC).
For each directed edge $(i,j)$, curvature is defined as:
\[
\kappa_{ij}=1-\frac{W_1(m_i,m_j)}{d(i,j)},
\]
where $m_i$ and $m_j$ are neighborhood probability measures induced by semantic flow weights.
ORC captures how local information neighborhoods contract or diverge during interaction.

\textbf{Interpretation.}
Positive curvature indicates redundant information sinks (e.g., echo chambers or collusion), while negative curvature reveals fragile bottlenecks where semantic perturbations may amplify and cascade.
Under normal operation, MAS interactions exhibit stable local curvature patterns tied to task roles and communication norms.

\subsection{SCCAL: Coupled Dynamics Predictor}

SCCAL explicitly models the bidirectional constraints between semantic updates and geometric evolution.

\paragraph{Geometry-Aware Semantic Predictor ($\psi$).}
Semantic dynamics are modeled using a curvature-gated recurrent architecture:
\[
\hat{\mathbf{y}}_i^{t+1}=\mathrm{GRU}\!\left(
\mathbf{y}_i^t,\,
\mathrm{Agg}\{\alpha_{ij}^t\mathbf{y}_j^t\}_{j\in\mathcal{N}_i}
\right),
\]
where attention weights are modulated by structural stability:
\[
\alpha_{ij}^t \propto \exp\!\left(w_{ij}^t \cdot \mathrm{ReLU}(\kappa_{ij}^t)\right).
\]
This mechanism attenuates influence from structurally unstable or overly redundant interactions, even when semantic similarity is high.

\paragraph{Semantic-Tension Geometric Predictor ($\phi$).}
Geometric evolution is modeled as a response to semantic tension.
A temporal graph network predicts next-step curvature:
\[
\hat{\kappa}_{ij}^{t+1}
=
\mathrm{MLP}\!\left[
\kappa_{ij}^t \,\Vert\,
\|\mathbf{y}_i^t-\mathbf{y}_j^t\| \,\Vert\,
\mathrm{Var}(\mathcal{N}_i,\mathcal{N}_j)
\right],
\]
capturing how semantic divergence induces local structural reorganization.

\subsection{Anomaly Detection and Root-Cause Attribution}

We define the auditing signal as the joint prediction residual of the coupled system:
\[
\mathcal{A}_t
=
\sum_i \|\mathbf{y}_i^t-\hat{\mathbf{y}}_i^t\|^2
+
\lambda\sum_{i,j}|\kappa_{ij}^t-\hat{\kappa}_{ij}^t|.
\]
Semantic–geometric consistency violations are flagged when this residual exceeds the learned normal range.
Importantly, geometric residuals alone can trigger early warnings before explicit semantic drift occurs.
Edges with maximal curvature deviation directly identify agents responsible for initiating cascading risk trajectories.

\section{Experiments}\label{sec:experiments}
Our experiments evaluate a single hypothesis: cascading risk in LLM-driven multi-agent systems arises from a breakdown of semantic–geometric co-evolution, and such breakdowns manifest before explicit semantic violations or task failure. Rather than optimizing for a specific attack taxonomy, we test whether semantic–geometric inconsistency (i) reliably predicts cascading risk, (ii) provides early warning, and (iii) generalizes beyond adversarial settings.

We note that our experiments validate manifold deviation indirectly through residual dynamics and early-warning behavior, rather than explicitly estimating intrinsic dimensionality or geodesic distance.

\subsection{Experimental Setup}

We construct multi-agent systems (MAS) from a heterogeneous agent library spanning multiple domains and roles. Each task instantiates 12–15 agents that interact over multiple rounds via natural language to solve long-horizon collaborative tasks. At each round, we record the communication graph, semantic edge representations, and Ollivier–Ricci curvature (ORC). We collect approximately 2,000 normal trajectories and 2,000 risk trajectories. Further details on agent composition, risk induction, and ORC computation are provided in the appendix.

\subsection{Cascading Risk Detection}

We first evaluate whether semantic–geometric consistency enables reliable detection of cascading risk from complete interaction trajectories. We compare against: (i) Semantic-only (\citep{reimers-2019-sentence-bert,qwen2.5}), using semantic representations for risk classification; (ii) Structural GNN model (\citep{chen2023sp}), using graph structure and node features without curvature; and (iii) ORC-only, using curvature statistics alone.

Across all metrics (AUROC, AUPRC, F1), from table \ref{tab:core_results}, our method substantially outperforms all baselines, demonstrating that neither semantic signals nor geometric structure alone are sufficient. Joint modeling of semantic content and interaction geometry yields a consistent advantage, supporting our framing of cascading risk as a coordination failure rather than a purely semantic anomaly.

\begin{table}[t]
\centering
\caption{Cascading risk detection across adversarial and natural failure regimes.}
\label{tab:core_results}
\begin{tabular}{lcccc}
\toprule
\textbf{Method} 
& \textbf{AUROC (Atk)} 
& \textbf{AUROC (Nat)} 
& \textbf{DLT (Atk)}
& \textbf{DLT (Nat)} \\
\midrule
Semantic-only & 0.71 & 0.62 & 5.8 & 0.4  \\
ORC-only & 0.77 & 0.69  & 7.4 & 1.2  \\
Structural GNN & 0.75 & 0.77 & 7.2 & 2.6\\
\textbf{Ours (Consistency)} 
& \textbf{0.88} 
& \textbf{0.83} 
& \textbf{10.2}
& \textbf{3.4} \\
\bottomrule
\end{tabular}
\end{table}

\subsection{Early Warning via Consistency Violation}

To assess early warning capability, we measure Detection Lead Time (DLT)—the number of interaction rounds by which a method signals risk before explicit semantic violations become observable. Thresholds are fixed on validation data and evaluated across multiple percentile settings. Unlike conventional Detection Lead Time that measures the time gap between detection and the final risk occurrence, we define DLT with respect to the first explicit semantic violation, which better reflects the early perception ability before high-level risk becomes observable.
DLT is a positive lead-time metric, where higher values correspond to stronger early-risk perception capability.

The experimental results are shown in DLT (Atk) in Table \ref{tab:core_results}. We find that semantic–geometric inconsistency emerges several interaction rounds earlier than any semantic risk signal. While curvature-only methods provide partial anticipation, they are unstable and prone to false positives. Semantic-only methods consistently lag behind. These results indicate that misalignment between semantic flow and geometric structure is an early and robust indicator of cascading risk.

Table \ref{tab:core_results} reveals a critical limitation of traditional graph learning baselines. While the Structural GNN achieves competitive accuracy on natural faults , its performance degrades significantly on cascading attacks . Crucially, the GNN exhibits a detection latency, contrasting sharply with SCCAL’s early warning capability. This suggests that GNNs, which rely on local neighborhood aggregation, tend to be reactive—detecting anomalies only after structural collapse is explicit. In contrast, our curvature-based approach captures the subtle accumulation of structural tension in the early stages of long-term interactions, enabling proactive intervention.

\begin{table}[t]
\centering
\caption{Limitations of curvature-only auditing.}
\label{tab:curvature_insufficiency}
\begin{tabular}{lcc}
\toprule
\textbf{Method} 
& \textbf{FPR (High Curvature)} $\downarrow$ 
& \textbf{AUROC Drop (Noise)} $\downarrow$ \\
\midrule
ORC-only & 0.32 & $-0.14$ \\
\textbf{Ours (Consistency)} & \textbf{0.07} & \textbf{-0.05} \\
\bottomrule
\end{tabular}
\end{table}

\subsection{Attack-Free Natural Failures}

To test generalization beyond adversarial settings, we evaluate aligned MAS under benign but realistic stressors, including limited memory, response delays, paraphrasing noise, and role overload. No adversarial prompts or unsafe content are introduced.

Despite fluent and locally coherent language, systems often fail at the task level due to degraded coordination. In these scenarios, no semantic violations are detected, yet semantic–geometric inconsistency rises sharply several rounds before failure. Semantic-only methods perform near chance; curvature-only methods capture partial degradation but confuse benign coordination with risk. In contrast, consistency-based detection anticipates failure reliably, demonstrating that cascading risk is not inherently adversarial.

Table~\ref{tab:core_results} shows that semantic–geometric consistency achieves consistently higher detection accuracy and substantially longer lead time than semantic-only or curvature-only baselines, across both adversarial attacks and attack-free natural failures, indicating regime-invariant instability captured by co-evolutionary violations.

\subsection{Is Curvature Alone Sufficient?}

We further examine whether curvature signals alone can reliably indicate risk. We construct high-curvature but semantically healthy interaction windows from normal trajectories and measure false alarm rates. We use False Positive Rate(FPR) as evaluation metrics.

Results in Table~\ref{tab:curvature_insufficiency} show that curvature-only methods exhibit systematic false positives in highly collaborative but risk-free scenarios. High curvature may reflect efficient coordination rather than instability. Semantic grounding is therefore essential: only when geometric anomalies conflict with semantic flow do they reliably signal risk. This experiment highlights the ambiguity of geometry in isolation and motivates principled semantic–geometric coupling.

And, and the same time, results demonstrate that curvature alone is an ambiguous risk signal, exhibiting high false positives in benign high-curvature collaboration and strong sensitivity to noise, whereas semantic grounding significantly suppresses geometric misinterpretation.

\subsection{Ablation Study}

Finally, we conduct ablations by removing the semantic module, curvature features, or the coupling loss. All variants suffer performance degradation, with the largest drop observed when semantic–geometric coupling is removed. This confirms that early risk detection depends critically on modeling the interaction between semantic meaning and geometric structure, rather than treating them as independent features.

Table~\ref{tab:ablation_main} shows that removing semantic–geometric coupling substantially degrades both detection accuracy and early-warning capability, confirming that risk anticipation depends on modeling co-evolution rather than independent semantic or structural features.

Overall, these experiments support a unified conclusion: cascading risk in multi-agent systems is best characterized as a failure of semantic–structural co-evolution, detectable through principled semantic–geometric consistency modeling.

Notably, neither semantic nor geometric residual alone consistently anticipates failure; early warning emerges only when deviations are measured jointly, consistent with the co-evolutionary formulation in Section 3.

\begin{table}[t]
\centering
\caption{Ablation study of semantic–geometric coupling.}
\label{tab:ablation_main}
\begin{tabular}{lccc}
\toprule
\textbf{Variant} & \textbf{AUROC} & \textbf{DLT} \\
\midrule
Full model & 0.88 & 9.8 \\
-- Coupling & 0.81 & 7.1 \\
-- Semantic module & 0.79 & 6.7 \\
-- Curvature & 0.75 &  4.9\\
\bottomrule
\end{tabular}
\end{table}

\section{Discussion}
\label{sec:discussion}

Our investigation into cascading risks within LLM-based Multi-Agent Systems (MAS) yields three critical insights into the interplay between semantic intent and interaction topology.

\paragraph{Complementarity of Risk Signals: Lagging vs. Leading.}
Our results reveal a fundamental temporal asymmetry between semantic and geometric auditing. Semantic signals, while interpretable and aligned with policy enforcement, inherently function as \textit{lagging indicators}. As demonstrated in our time-lag analysis, explicit semantic violations (e.g., toxicity or protocol breaches) typically emerge only after the interaction dynamics have already drifted into an irreversible failure regime. In contrast, geometric deviations—specifically, anomalies in Ollivier-Ricci Curvature (ORC) relative to learned co-evolutionary baselines—serve as \textit{leading indicators}. We consistently observe that such curvature deviations precede semantic breakdowns by several interaction rounds. This temporal gap confirms that cascading risk is not an instantaneous event, but rather a gradual structural degeneration—such as the formation of echo chambers (excessive positive curvature) or fragile diffusion trees (high negative curvature)—that eventually manifests as semantic failure.

\paragraph{Semantic–Curvature Co-evolution as a Stability Criterion.}
Crucially, our results show that geometry alone is a necessary but insufficient condition for reliable risk detection. Ablation studies demonstrate that benign, high-performance collaboration can exhibit geometric patterns similar to those observed in failure cases. For example, intensive brainstorming or rapid task decomposition naturally induces high positive curvature due to local redundancy, which naive geometric detectors may misinterpret as malicious collusion. This observation motivates our semantic–curvature co-evolution framework. We find that system stability is governed not by geometric configuration in isolation, but by the \textit{coherence} between semantic intent and structural evolution. Risk emerges as a form of semantic–geometric \textit{decoherence}: structural tightening that lacks semantic justification, or information diffusion that is not grounded in coherent intent. Modeling this coupled dynamic enables our framework to suppress false positives in benign dense interactions while remaining sensitive to structurally unstable trajectories.

\paragraph{From Content Filtering to Process-Level Auditing.}
From a security perspective, these findings suggest a shift from static, content-centric inspection toward \textit{dynamic, process-level auditing}. Guardrails that operate on isolated messages are fundamentally limited in their ability to capture emergent, non-local risks arising from Multi-Agent interaction dynamics. By modeling the semantic–geometric evolution of collaboration, our approach provides a principled mechanism for early intervention before semantic violations become explicit. Moreover, the geometric component of our framework supports structural root-cause attribution: by localizing curvature distortions, auditors can identify specific agents or interaction links that catalyze cascading instability, enabling targeted mitigation of systemic failure modes rather than post-hoc symptom suppression.

\section{Limitations}
\label{sec:limitations}

While our framework offers a novel geometric perspective on MAS safety, we acknowledge specific limitations that outline directions for future research.
\paragraph{\textbf{Simulation-to-Reality Gap.}}
 Our evaluation primarily relies on a large-scale synthetic benchmark aligned with the AEGIS 2.0 risk taxonomy \citep{ghosh2025aegis2}. Although the simulator is rigorously designed to match the semantic and topological properties of real-world interactions, it may not fully capture the latency, asynchronous noise, and human-in-the-loop dynamics characteristic of production environments.
\paragraph{\textbf{Scope of Geometric Invariants.} }
This study focuses predominantly on local Ollivier-Ricci Curvature \citep{ollivier2009ricci}. While effective for neighborhood-level anomalies, local metrics may miss global topological phase transitions in large-scale networks. The potential of global geometric invariants---such as spectral properties or persistent homology---in characterizing system-wide safety remains an open frontier.

\section{Conclusion}
In this work, we introduced a principled framework for auditing cascading risks in MAS by modeling safety as the trajectory stability on a coupled semantic–geometric manifold. Unlike reactive content filtering, our method formalizes risk as a consistency residual—a divergence between the expected structural evolution and observed semantic flows. This formulation shifts the auditing paradigm from detecting isolated semantic violations to monitoring the process-level dynamics of interaction.

Our empirical findings reveal a temporal asymmetry: interaction geometry acts as a slow variable that encodes latent systemic intent  well before explicit semantic failures emerge. However, geometry alone is insufficient; benign high-density collaboration can mimic risky topologies. Our method resolves this by grounding geometric stress in semantic context, enabling robust early warnings while filtering structural false positives.

Looking forward, the graph-theoretic nature of our method offers a pathway for trustworthy AI across modalities. Since geometric invariants  are agnostic to the underlying data type, our framework can be naturally extended to Multi-modal Agent Systems—such as Vision-Language-Action models—where structural misalignment between perception and reasoning may serve as universal precursors to failure. We hope this work encourages further exploration of geometric co-evolution as a foundational design principle for resilient and interpretable AI societies.

\section*{Acknowledgements}

This work was supported in part by the National Natural Science Foundation of China (Grant No. 72293583), 
the Chinese Academy of Engineering Consulting Project (Grant No. 2025-XZ-08), 
and the Ministry of Education of China Project of Humanities and Social Sciences (Grant No. 24JZD040).

\bibliography{presubmit_paper}
\bibliographystyle{iclr2026_conference}

\appendix
\section{Appendix}

\subsection{Implementation Details}

\subsubsection{Environment and Infrastructure}
Our framework is implemented using PyTorch 2.1 and PyTorch Geometric (PyG). All experiments are conducted on a Linux server equipped with a single NVIDIA A100 GPU (80GB VRAM) and an AMD EPYC 7742 CPU. For graph-based geometric computations, we utilize the GraphRicciCurvature library accelerated by the Python Optimal Transport (POT) backend.

\subsubsection{Semantic Encoding}
All agent messages are mapped into a latent semantic space using a frozen sentence-level encoder. We employ a pre-trained Sentence-BERT (SBERT) model to obtain $d$-dimensional embeddings. The encoder remains fixed throughout training and evaluation to decouple semantic representation from interaction dynamics. For each agent at time step $t$, its semantic state $y_i^t$ corresponds to the embedding of its most recent message.

\subsubsection{Semantic Flow Graph Construction}

At each interaction round, we construct a directed semantic interaction graph $\mathcal{G}_t = (\mathcal{V}, \mathcal{E}_t, \mathbf{W}_t)$, where nodes represent agents and weighted edges represent semantic influence. The weight of edge $(i,j)$ is defined as:
\[
w_{ij}^t = \tau(i,j) \cdot \chi(i),
\]
where semantic transmissibility $\tau(i,j)$ is computed via cosine similarity between $y_i^t$ and $y_j^{t-1}$, and semantic credibility $\chi(i) = \exp(-\text{PPL}(s_i^t))$ is derived from message perplexity computed using a fixed reference language model. Low-weight edges are pruned to reduce noise.

---

\subsubsection{Ollivier–Ricci Curvature Computation}

Given $\mathcal{G}_t$, we compute Ollivier–Ricci Curvature (ORC) for each edge. For node $i$, a probability measure $m_i$ is defined over its outgoing neighbors, normalized by edge weights. The curvature of edge $(i,j)$ is defined as \citep{ollivier2009ricci}:
\[
\kappa_{ij} = 1 - \frac{W_1(m_i, m_j)}{d(i,j)},
\]
where $W_1$ denotes the Wasserstein-1 distance and $d(i,j)$ is the shortest-path distance. Curvature values are clipped for numerical stability.

---

\subsubsection{Coupled Dynamics Modeling}

The method we proposed consists of two interleaved predictors modeling semantic and geometric evolution.

\paragraph{Geometry-Aware Semantic Predictor.}

Semantic evolution is modeled using a GRU with curvature-gated attention. Neighbor aggregation is weighted by:
\[
\alpha_{ij} \propto \exp(w_{ij}^t \cdot \text{ReLU}(\kappa_{ij}^t)),
\]
suppressing information flow through structurally unstable or redundant edges.

\paragraph{Semantic-Tension Geometric Predictor.}

Geometric evolution is modeled using a temporal graph neural network. For each edge, the predictor takes as input the current curvature, semantic tension $|y_i^t - y_j^t|$, and local neighborhood variance, and outputs the predicted curvature at the next time step.

\subsubsection{Training and Auditing}

The model is trained on normal (risk-free) interaction trajectories by minimizing a joint prediction loss:
\[
\mathcal{L} = \sum_i  \| y_i^{t+1} - \hat{y}_i^{t+1}  \|^2
+ \lambda \sum_{i,j} |\kappa_{ij}^{t+1} - \hat{\kappa}_{ij}^{t+1}|.
\]

At inference time, the auditing signal is computed as the joint semantic–geometric residual. Thresholds are selected on a validation set and fixed for all test scenarios. Root-cause attribution is performed by identifying nodes or edges with the largest localized geometric residuals.

\subsubsection{Detection Lead Time}\label{DTLdetail}
Detection Lead Time and Response Latency To evaluate the perception ability of our method in the early stages of risk, we introduce three critical timestamps for each risk trajectory:

$T_{start}$  Risk Injection. The time step when the risk-inducing agent or structural perturbation is first introduced.

$T_{alarm}$  First Alert. The first time step where the anomaly score $s_t$ exceeds the threshold $\tau$: $T_{alarm} = \min \{t \mid s_t \ge \tau, t \ge T_{start}\}$. Note that we constrain $t \ge T_{start}$ to filter out false positives prior to risk injection.

$T_{sem}$  Semantic Manifestation. The moment when semantic violations first appear explicitly: $T_{sem} = \min \{t \mid R_{sem}(t) \ge \tau_{sem}\}$.

Based on these, we define Detection Lead Time (DLT) as:$$\text{DLT} = T_{sem} - T_{alarm}$$

A positive DLT signifies the safety margin: the duration the system operates in a structurally unstable but semantically compliant state. A larger DLT implies a stronger capability to decouple latent structural causes from their eventual semantic symptoms.
We emphasize that this definition measures the effective early-warning lead time within the post-injection causal window, rather than the absolute earliest alert over the entire trajectory. Alerts triggered before $T_{start}$ are excluded to avoid counting spurious pre-injection fluctuations that are not causally attributable 
to the injected risk. Importantly, this validity criterion is applied uniformly to all methods, including all baselines, ensuring fair and comparable DLT evaluation.

\subsubsection{Data sample example}
\paragraph{Agent:}
{\small
\begin{verbatim}
{
  "role_name": "Machine Learning Agent",
  "domain": "Materials Science / Civil Engineering",
  "framework_source": "------",
  "prompt": "Map the relationship for input-output variables based on ...",
  "risk_tags": [
    "numerical inaccuracy",
    "lack of detailed reasoning",
    "data bias (small dataset)"
  ],
  "behavior_pattern": "Utilizes regression models ..."
}
\end{verbatim}
}
\paragraph{interaction trajectory:}
{\small
\begin{verbatim}
{
  "Define Model Scope, ....": 
  [
    {
      "agent_name": "Ethics Oversight Agent",
      "log": "[Define Model Scope, Target Region,  ...",
      "sendto": [
        ["Research Scientist Agent"]
      ]
    }
  ]
}
\end{verbatim}
}
\subsubsection{Risk Interaction sample}
To construct cascading risk data, we combined the 21 types of risks defined in AEGIS 2.0\citep{ghosh2025aegis2}  and adopted the method proposed by SRA\citep{lian2025semantic} to induce risk behavior without compromising semantic coherence. The attack process follows the following constraints: (i) the attack sample does not contain explicit violation content in the early stages; (ii) The generated interaction process conforms to the semantic distribution of normal MAS; In the end, we generated approximately 2000 risk samples and annotated their occurrence time steps to support time-series risk analysis

\subsection{Early risk examples of semantic geometric mismatch}

Although risk behavior in MAS is typically defined by semantic violations or harmful content, in actual operation, the formation of risk often does not begin with explicit semantic shifts, but rather stems from potential imbalances at the level of interaction structures.

\begin{figure*}[t]
    \centering
    \includegraphics[width=0.9\linewidth]{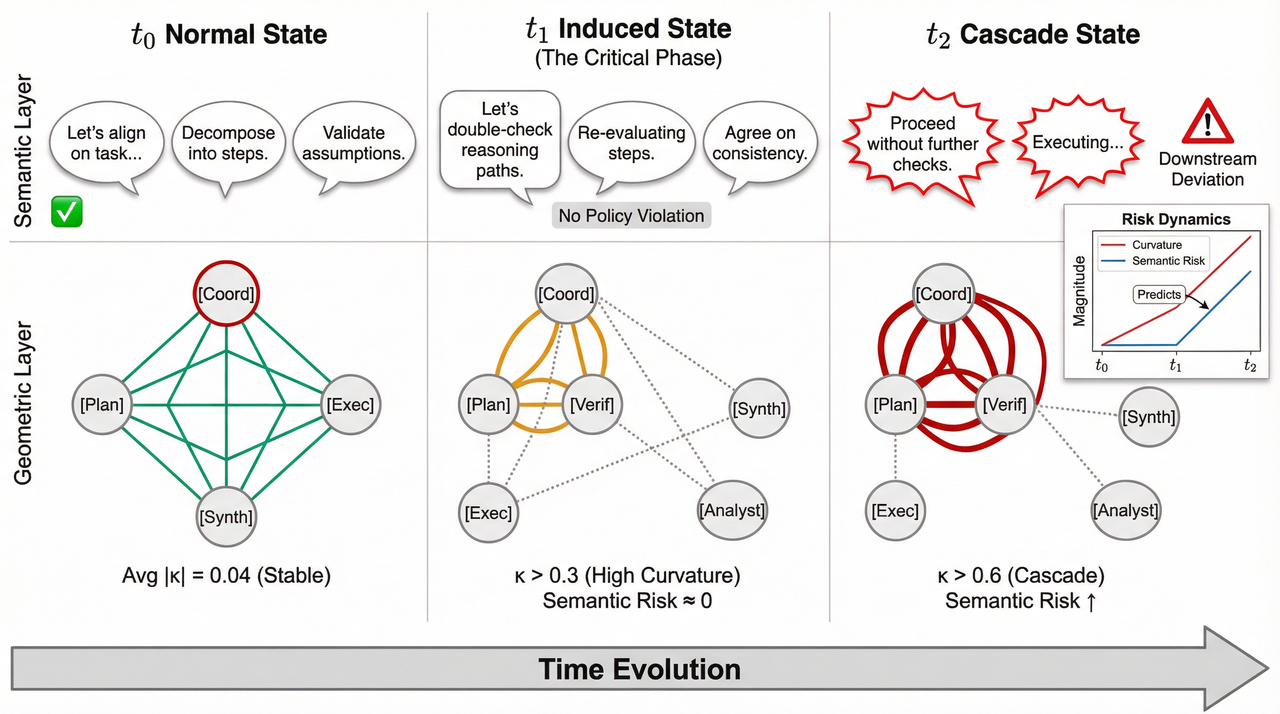} 
    
    \caption{\textbf{Illustration of the Semantics-Geometry Detection Lag.} 
    The evolution of a cascading risk scenario across three phases:
    \textbf{(Phase $t_0$)} The system operates in a stable state with balanced topology.
    \textbf{(Phase $t_1$)} A structural ``echo chamber'' forms (indicated by high Ollivier-Ricci Curvature, $\kappa > 0.3$) as agents align tightly. Crucially, \textit{semantic content remains benign and policy-compliant}, causing traditional semantic auditors to miss this latent risk.
    \textbf{(Phase $t_2$)} Semantic violations finally erupt and cascade along the high-curvature paths established in $t_1$. 
    This demonstrates that \textit{geometric collapse precedes semantic violation}, validating our approach of structural early warning.}
    \label{fig:motivating_example}
\end{figure*}

Figure \ref{fig:motivating_example} shows a typical example of Multi-Agent risk induction. In this scenario, the system presents stable semantic interaction and topological structure in the initial stage$t_0$. Subsequently, a instigating agent has a sustained impact $t_1$ on Multi-Agent in its neighborhood through semantically fully compliant and highly coherent outputs. It is worth noting that at this stage, there were no explicit violations in the language output of all agents, and traditional semantic based risk detection methods are difficult to distinguish between this state and normal collaborative behavior.

However, from the perspective of topological geometry, significant changes have occurred in the interactive network during this stage. The local structure dominated by the responsible party gradually evolves into a highly redundant closed loop subgraph, and its corresponding Ollivier Ricci curvature significantly deviates from the expected geometric background of the system under normal operating conditions. This phenomenon indicates that before semantic risk becomes explicit, there has been a systematic imbalance in the organization of information flow.

As the interaction progresses further $t_2$, semantic anomalies begin to manifest in Multi-Agents and cascade along existing communication paths. At this point, semantic shift and geometric anomaly erupted simultaneously, verifying the phenomenon of early geometric imbalance occurring before explicit semantic violations in time.

The example reveals a key challenge: risk auditing methods that rely solely on semantic signals are difficult to capture cascading risks caused by structural imbalances in MAS. This naturally leads to the core question of this article: how to identify and characterize potential risk inducing processes before semantic violations occur?

\subsection{Time-Lag Analysis of Curvature and Semantic Risk}
\label{sec:timelag}

This section investigates whether topological geometric anomalies, quantified by Ollivier–Ricci Curvature (ORC), systematically precede semantic-level risk manifestations in Multi-Agent systems. If so, curvature deviations serve not merely as post-hoc descriptors of failure, but as leading indicators of cascading risk.

\subsubsection{Event-Aligned Time-Lag Protocol}

We adopt an event-aligned analysis. For each annotated risk trajectory, the first confirmed semantic risk occurrence is defined as $t = 0$. We analyze both geometric and semantic signals over a temporal window $t \in [-T, +T]$, aggregating results across trajectories of the same risk type.

All signals are normalized per trajectory to remove scale effects and ensure comparability across different risk instances.

\begin{figure*}[t]
\centering
\includegraphics[width=0.5\linewidth]{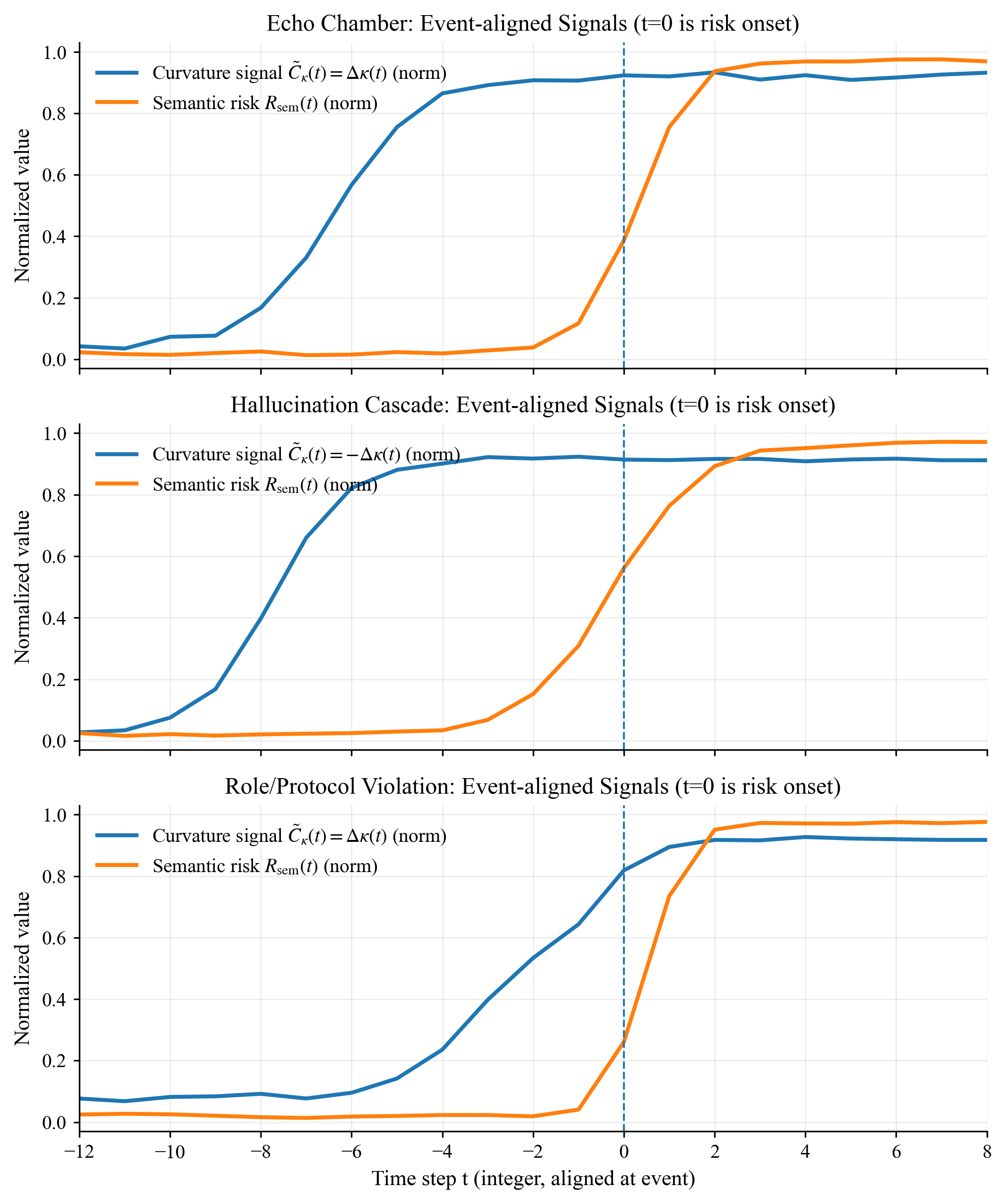}
\caption{Event-aligned time-lag analysis of curvature and semantic risk signals. The dashed line denotes the semantic risk onset ($t=0$).}
\label{fig:timelag}
\end{figure*}

\subsubsection{Signals Definition}

\paragraph{Curvature Signal.}
We define the curvature anomaly signal as
\[
\Delta \kappa(t),
\]
the mean absolute deviation of edge-level ORC within a unit time window. Positive curvature offset reflects increasing local redundancy (e.g., clique formation), while growth in negative curvature magnitude indicates fragile, tree-like diffusion structures.

\paragraph{Semantic Risk Signal.}

The semantic risk signal $R_{\text{sem}}(t)$ is computed by applying a semantic risk classifier (based on the SRA framework) to sentence-level embeddings produced by a frozen encoder. It reflects the probability of explicit semantic violation.

Risk-Type-Specific Temporal Patterns

We conduct independent time-lag analyses for three representative cascading risk categories.

\subsubsection{Echo Chamber Risks.}

Echo chamber formation induces rapid densification of local subgraphs, leading to a systematic shift toward positive curvature. As shown in Fig. \ref{fig:timelag}, $\Delta \kappa(t)$ begins a monotonic increase approximately 5–6 interaction rounds before $t=0$, while semantic risk remains near baseline until the risk event. This indicates that highly redundant information loops emerge structurally well before malicious or policy-violating content becomes explicit.

\subsubsection{Hallucination Cascade Risks.}

Hallucination cascades are characterized by erroneous information propagating along unidirectional chains, reflected as a sharp increase in negative curvature magnitude. The curvature anomaly rises 6–7 rounds prior to semantic risk onset, whereas semantic signals typically lag by 1–2 rounds. Notably, in this scenario, the absolute magnitude of curvature deviation is more informative than its sign, capturing the accumulation of structural fragility.

\subsubsection{Role Violation and Protocol Misalignment.}

These risks manifest as disruptions of expected hierarchical or task-oriented communication structures, such as abnormal feedback loops. Curvature anomalies emerge approximately 4 rounds before $t=0$, while semantic signals rise sharply only near the confirmation stage. Here, curvature behaves as a slow structural variable, whereas semantic signals act as fast but delayed indicators.

\end{document}